\documentclass[a4paper,11pt]{article}
\usepackage[commandnameprefix=always]{changes}
\usepackage[T1]{fontenc}
\usepackage{lmodern,graphicx,booktabs}
\usepackage{comment}
\usepackage{mathtools}
\usepackage{multirow}
\usepackage{hyperref}
\usepackage{cleveref}
\usepackage{changes,xcolor,subcaption}
\usepackage{url}
\usepackage{siunitx} 
\usepackage{parskip} 
\usepackage[sort&compress,numbers,comma,square,super]{natbib}

\usepackage[top=1in,bottom=1in,left=1in,right=1in]{geometry}
\begin{document}
{\centering{ \textbf{Design of radiative cooling paint coating and insights into its sub-ambient cooling behaviour}\par}}
{\centering{Bhrigu Rishi Mishra$^*$, Sreerag Sundaram$^*$, Karthik Sasihithlu\\
$^*$ joint first authors
\par}}
{\centering{Department of Energy Science and Engineering, Indian Institute of Technology Bombay, Powai, Mumbai 400076,
Maharashtra, India\par}
}
\section*{Keywords}
Radiative cooling paint, passive daytime radiative cooling, cooling performance measurement, lumerical FDTD, Kubelka Munk theory
\section*{Abstract}
Recent developments in radiative cooling technologies have primarily focused on affordable paint coatings that are easy to fabricate and deploy. Using a systematic approach to obtain optimal parameters, a radiative cooling (RC) paint coating using titanium dioxide (TiO$_2$) and polydimethylsiloxane (PDMS) is designed. The resulting paint exhibits a high solar reflectivity of 88.2~\% (more than 94\% in visible and NIR) and an emissivity of 92.4~\%. Outdoor testing demonstrates a maximum reduction of 7.9~$^\circ$C in the internal temperature of an RC paint-coated aluminium (Al) box compared to a bare Al box but in contrast to other studies, no sub-ambient cooling have been observed. 
In this context, a comprehensive analysis explaining the absence of sub-ambient cooling and underscore the importance of a standardized reporting methodology for RC paints has been discussed. Theoretical calculations suggest that the developed RC paint can achieve sub-ambient cooling (1-4 $^\circ{}$C) under specific ambient conditions.

\section{Introduction}
\label{sec:intro}

As the global climate crisis continues to escalate, there is an urgent need for innovative and sustainable solutions to mitigate the adverse effects of rising temperatures and reduce energy consumption. In this pursuit, radiative cooling coatings have emerged as a promising avenue for enhancing passive cooling strategies in various applications, ranging from buildings and automobiles to electronic devices. These specialized coatings offer a sustainable, clean, and electricity-free cooling solution to counteract the urban heat island effect and reduce the dependence on energy-intensive and polluting air conditioning systems. Apart from effectively reflecting incoming solar radiation, these coatings are engineered to release thermal radiation in the long-wavelength infrared (LWIR) spectrum. This LWIR radiation has the unique ability to pass through the Earth's atmosphere via the transparency window, which spans the 8--13~$\mu$m wavelength range, and subsequently disperse into the cold depths of outer space. This phenomenon, known as passive radiative cooling, allows surfaces coated with such coatings to cool down significantly below ambient temperatures, even in the midst of intense sunlight. Theoretically, it can potentially reduce the temperature of a surface, passively, by 60~$^\circ$C \cite{Chen2016}. 

Different types of structures, such as multilayer \cite{Raman2014}, glass-polymer hybrid metamaterial \cite{Zhai2017}, biologically inspired structures \cite{zhang2020biologically,liu2021biomimetic},  hierarchical porous polymer \cite{Mandal2018}, and disordered coatings \cite{Atiganyanun2018,Li2020CaCO3,Li2021baso4,Zhang2021Zro2,Xue2020,zhang2023efficient} have been explored aiming to achieve the desired radiative properties in respective spectral bands. 
Among these structures, disordered coatings or paint coatings have gained a lot of popularity due to its simple design, ease of fabrication and low cost. 
 
Materials such as TiO$_2$ \cite{Mishra2021}, SiO$_2$ \cite{Atiganyanun2018}, CaCO$_3$ \cite{Li2020CaCO3}, BaSO$_4$ \cite{Li2021baso4}, ZrO$_2$ \cite{Zhang2021Zro2}, and Al$_2$O$_3$ \cite{zhang2023efficient}, have been used to achieve high solar reflection. Among these, high band gap materials such as SiO$_2$, BaSO$_4$, CaCO$_3$, Al$_2$O$_3$ require large volume fractions (>~50~\%) to counterbalance their low scattering on account of the small difference in refractive indices of these pigments and polymers (0.05-0.25). A high pigment volume concentration in a paint increases the cost and the consequent decrease in polymer (binder) concentration also negatively affects the paint properties viz. its stress/strain tolerance, water resistivity, durability \cite{rooney2018effect}. Beyond a certain pigment concentration, a paint becomes fragile, less ductile, low in adhesive strength due to the development of internal stresses \cite{rodriguez2004influence, kasyanenko2018effect}. Therefore, these compounds are generally preferred to be used, by the paint industry, as auxiliary pigments -- to increase filler concentration of a paint composition, striking a balance between optimal paint properties and economics. 

In contrast to such materials, we use TiO$_2$ to develop a paint coating which can demonstrate passive daytime radiative cooling. Due to its high relative refractive index, it provides high reflectivity in visible and NIR with lesser volume fraction and thus provide longevity and durability compared to coatings with high band gap materials but lesser relative refractive index. We study and optimise the pigment particle size, its volume fraction, and the coating thickness to attain high solar reflectivity ($R_{\rm{solar}}$), aided by our novel, highly efficient semi-analytical technique discussed in a previous publication \cite{mishra2023semi}. 
Previous experimental research on coatings where TiO$_2$ is used as their sole pigment, have reported $R_{\rm{solar}}$ around 88--90~\% \cite{Bao2017, Xue2020, du2022daytime} with the reduction in net reflectivity brought about due to its highly absorbing nature in the solar spectrum's ultraviolet (UV) regime ($0.3-0.4~\mu$m). Mandal et al. \cite{mandal2020paints} have shown that TiO$_2$-based paints can achieve a solar reflectivity of 92--93~\% albeit with very high composition of the pigment by weight (93~wt.\%). In our previous work \cite{Mishra2021}, it was shown that theoretically it is possible to achieve $R_{\rm{solar}}$~=~94~\% in TiO$_2$-based coating, with just 15~\% volume fraction of the pigment.

In literature, the cooling performance of coatings for daytime RC is characterized by comparing the temperature of the coating to the ambient temperature during daytime. 
In fact, in outdoor tests, these designs report cooling by several degrees Celsius below ambient. The sub-ambient cooling is not only dependent on the radiative properties of the coating, and the prevelant environmental conditions, but also on the method adopted to characterise it. In most of the cases, the following practices have been adopted to report sub-ambient cooling:
\begin{itemize}
    \item  A radiation shield box, open at the top, has been used increase the coating's sky view factor  \cite{Mandal2018}.
    \item Polyethylene (PE) cover has been used to cover the open side of the radiation shield box to eliminate non-radiative losses \cite{Raman2014, Li2021baso4, du2022daytime}.
    \item Surface temperature of the coating is compared with the ambient temperature \cite{Mandal2018, Zhang2021Zro2, bijarniya2022experimentally, dasultra}.
    \item Ambient temperature is measured by keeping the temperature sensor/thermocouple  exposed to solar radiation and wind \cite{Mandal2018, Kou2017, Wang2021, Zhang2021Zro2, du2022daytime}.
\end{itemize}
These practices can sometimes artificially elevate the recorded readings. For instance, PE covers used to eliminate non-radiative losses and thus provide a better representation of radiative cooling performance, have been observed to trap heat inside the radiation shield box, leading to air temperatures measuring as high as 60-70~$^\circ$C \cite{Zhang2021Zro2,xia2023water}. This increase in air temperature provides large sub-ambient cooling compared to similar measurements done without PE cover \cite{xia2023water}. Moreover, the efficiency of these paints is also often contingent on environmental factors, such as humidity and atmospheric conditions, which can impact their cooling performance \cite{bijarniya2020environmental}. 

In the following sections, we first discuss the design and performance characteristics of our coating, followed by a discussion on the challenges faced by the radiative cooling community in reporting investigation results. 
 We analyze the impact of humidity and review current practices in reporting experimental tests, including the use of radiation shields and polyethylene covers to minimize convective losses, and the measurement of ambient temperature. Two very recent publications \cite{zhou2023best,bu2023consistent} have also highlighted these issues. We expand on these works with a goal o promote a unified methodology for reporting more accurate and comprehensive performance readings of novel radiative cooling paints, enabling easier comparison regardless of external conditions and human error.


\section{Design, material and methods}
\subsection{Design of coating}

\begin{figure}[t]
     \centering
     \begin{subfigure}[b]{0.46\textwidth}
    \centering
    \includegraphics[width=85 mm]{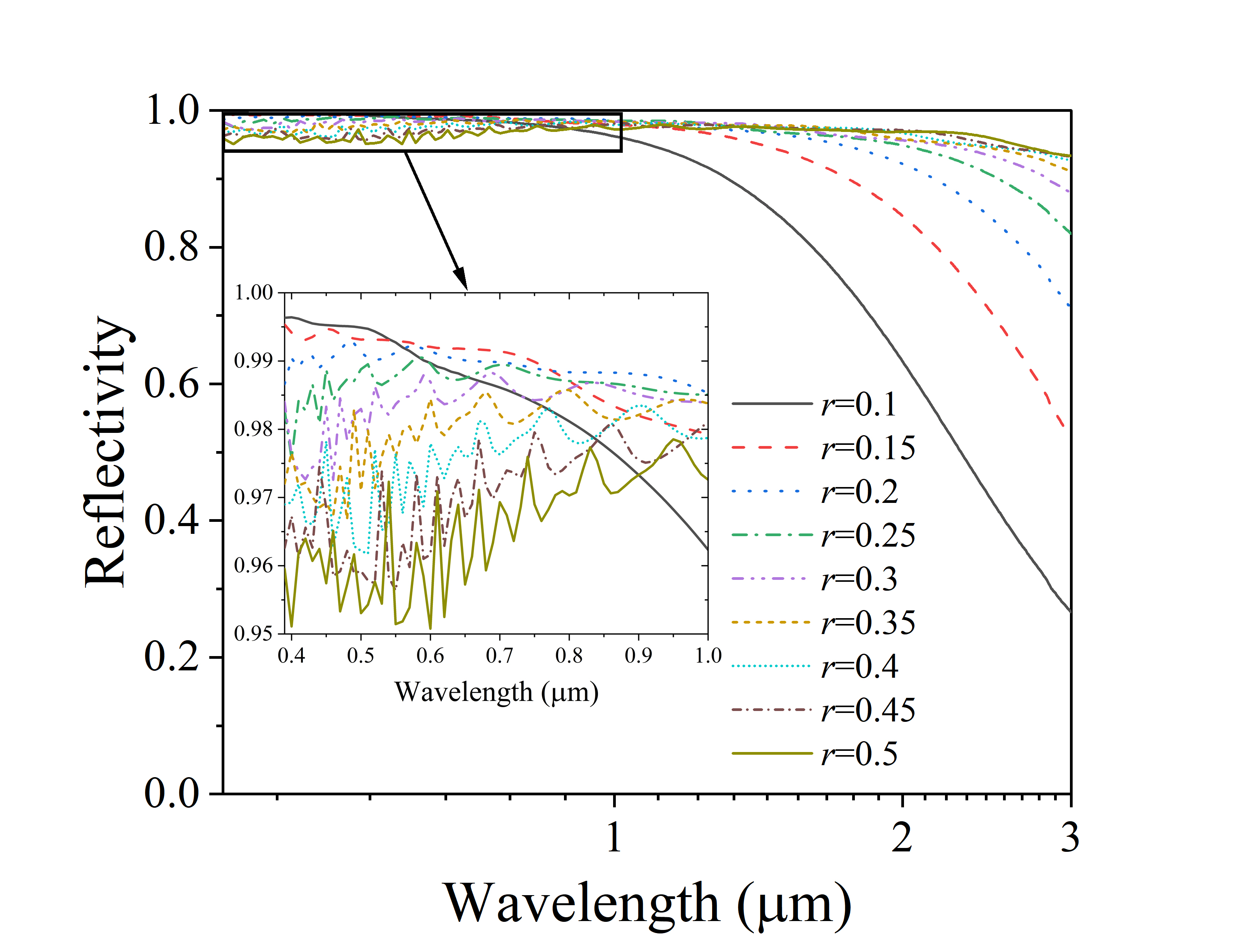}
    \caption{}
    \label{fig:R_for_r}
     \end{subfigure}
     \hfill
     \begin{subfigure}[b]{0.46\textwidth}
    \centering
    \includegraphics[width=85 mm]{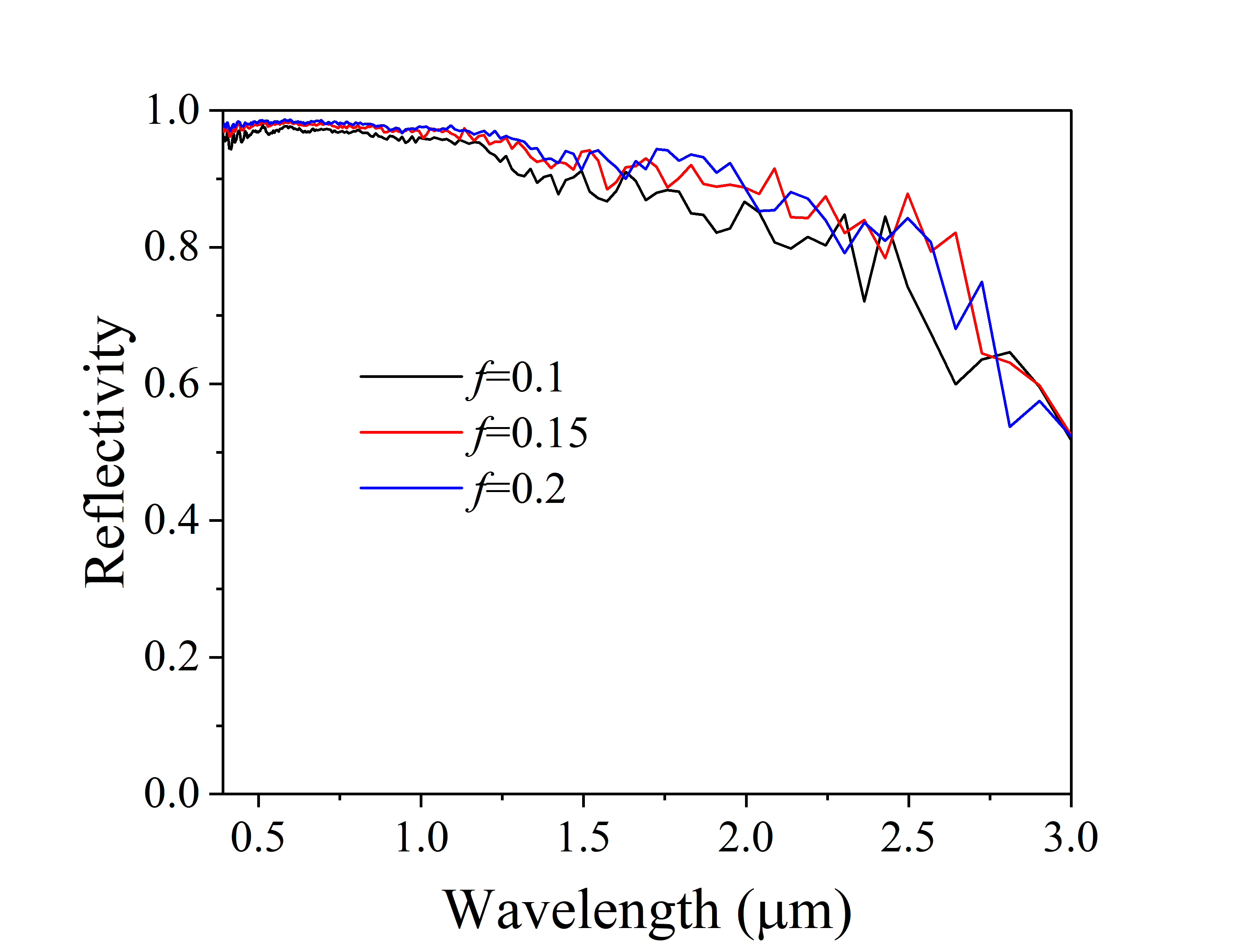}
    \caption{}
    \label{fig:R_for_f}
     \end{subfigure}
    \caption{Reflectivity of TiO$_2$/PDMS coating for (a) various radius ($r$) of particles, keeping $f$ (=5\%) and $L$ (=1~mm) constant, and (b) various volume fractions keeping $r$ (=0.25~$\mu$m) and $L$ (=300~$\mu$m) constant.}
    \label{fig:design_paint}
\end{figure}

The design of a disordered coating involves optimizing several parameters, including the particle's radius and volume fraction, as well as the coating thickness. To model the radiative properties of the coating, we rely on the Kubelka-Munk (KM) theory and a semi-analytical technique (detailed further in supplementary notes). The design process proceeds in two stages:
\begin{itemize}
 \item The KM theory \cite{Kubelka1947,Quinten2001,Molenaar1999} is first employed to determine the optimal radius for the spherical TiO$_2$ pigment particles.
\item
With the optimized pigment particle size as a given input, the pigment volume fraction and the coating thickness are fine-tuned using our proprietary semi-analytical technique \cite{mishra2023semi}.
\end{itemize}
In the first stage
the pigment volume fraction and coating thickness is kept constant at 5\% and 1 mm, respectively. This low volume fraction avoids any dependent scattering, and the larger thickness ensures that reflective performance isn't affected by variations in its depth. Since TiO$_2$ absorbs in the ultraviolet range up to 0.39~$\mu$m, all reflectance calculations aimed at optimizing design parameters are carried out over the 0.4--3~$\mu$m spectrum. The complex refractive indices of TiO$_2$ and PDMS used in the simulation are given in supplementary in Figure~S3. Solar reflectivity for various radii of TiO$_2$ particles is shown in \Cref{fig:R_for_r}. 
A notable observation from the inset of \Cref{fig:R_for_r} is the decline in spectral solar reflectance in the 0.4--0.9~$\mu$m band as particle radius increases. This is attributed to the red-shifting of the scattering resonance peak with increasing particle size. Subsequently, for larger particles, the reflectance curve flattens at longer wavelengths. The weighted solar reflectivity $R_{\rm{solar}} =\int I_{\text{AM1.5}}(\lambda) R(\lambda) d \lambda / \int I_{\text{AM1.5}}(\lambda) \, d\lambda$, where $I_{\text{AM1.5}}(\lambda)$ is the spectral solar irradiance \cite{AM1.5} over the solar spectrum ($\int I_{\text{AM1.5}}(\lambda) \, d\lambda$ = 986.5~W/m$^2$) -- is calculated for all reflectivity curves given in \Cref{fig:R_for_r} and tabulated in \Cref{table:R_for_r}. The $R_{\rm{solar}}$ increases with increasing pigment particle size until $r = 0.25~\mu$m, after which it decreases. This is because, with increasing $r$, the dip in spectral reflectivity (\Cref{fig:R_for_r}) between 0.4--0.9~$\mu$m outweighs the increase in reflectance in the NIR as the majority of incident solar energy is concentrated in the 0.3--1.8~$\mu$m band. 
As a result, a coating with pigment particle radii range 0.15~$\mu$m to 0.25~$\mu$m can achieve $R_{\rm{solar}}$ values comparable to those with a much larger radius, even though their spectral reflectivity is lower in the 2--3~$\mu$m range.

In the subsequent stage, we employ the semi-analytical methodology detailed in \cite{mishra2023semi} to determine the optimal volume fraction $f$ of the scattering pigment and the thickness of the RC coating. An advantage of this technique is that the calculations factor in the effect of dependent scattering by adjacent particles. 
As highlighted in \cite{mishra2023semi}, our semi-analytical approach begins by using exact numerical simulations to extract the scattering and absorption coefficients of a thinner layer. This data is then utilized with analytical tools, such as the Kubelka-Munk theory, to characterize layers with considerably larger thicknesses. Accordingly,  Ansys Lumerical FDTD \cite{Lumerical} is used to simulate and analyse the optical behaviour (reflectance and transmittance) for a 10-$\mu$m--thick coating at the following volume fractions: 10~\%, 15~\%, and 20~\%, with particle radius 0.25~$\mu$m (selected based on the previous analysis). KM theory is then used, with surface reflection corrections, to calculate the reflectivity of the TiO$_2$/PDMS coating for 300~$\mu$m thick layer.   
The results are presented in 
\Cref{fig:R_for_f} which shows the calculated reflectivity of the coating for varying $f$. 
Coatings with $f = 15$~\% and 20~\% show similar performance with $R_{\rm{solar}}$ = 96.5~\% and 97.0~\%, respectively. Thus we adopt 15~\% pigment volume fraction for experimental validation of the coating. 
Moreover, when examining the 15~\% volume fraction coating across different thicknesses, it becomes evident that beyond the 250-300~$\mu$m mark, any increment in reflectivity is marginal as seen in \Cref{fig:R_for_diff_L}.

\begin{table}[bh]
\caption{Weighted reflectivity of TiO$_2$/PDMS coating in wavelength range 0.4-3~$\mu$m calculated for various TiO$_2$ particle sizes with $f$ = 5\%, and $L$ = 1~mm.  }
\begin{tabular}{llllllllll}
\hline
Radius ($r$)       & 0.1    & 0.15   & 0.2    & 0.25   & 0.3    & 0.35   & 0.4    & 0.45   & 0.5    \\
Reflectivity ($R_{\rm{solar}}$) & 0.9495 & 0.9755 & 0.9822 & 0.9829 & 0.9812 & 0.9785 & 0.9750 & 0.9709 & 0.9665 \\ \hline
\end{tabular}
\label{table:R_for_r}
\end{table}

\subsection{Materials and methods}
We converted the 15~\% volume-based pigment/filler composition to a weight-based formulation. Mixing 70~\% PDMS (Part A, SYLGARD\textsuperscript{TM} 184, Dow) with o-Xylene (98~\%, Loba Chemie Pvt. Ltd.) in a 1:1 ratio, and stirring at 200 rpm gave a clear solution. We then added the TiO$_2$ pigment (Titanium (IV) oxide, rutile, Sigma-Aldrich) under slow stirring, and then increased stirring to 500 rpm for 30 minutes. The remaining PDMS and xylene, in the same ratio,  was subsequently added and stirred for 90 minutes. After resting the solution overnight, we applied two coats to aluminum substrates, allowing a 24-hour curing period at room temperature. We observe that xylene can be added at any stage to adjust viscosity without affecting optical properties.


\section{Results and discussion}
\subsection{Characterization of the coating}

\begin{figure}[t] 
    \centering
     \begin{subfigure}[b]{0.46\textwidth}
    \centering
    \includegraphics[width=85 mm]{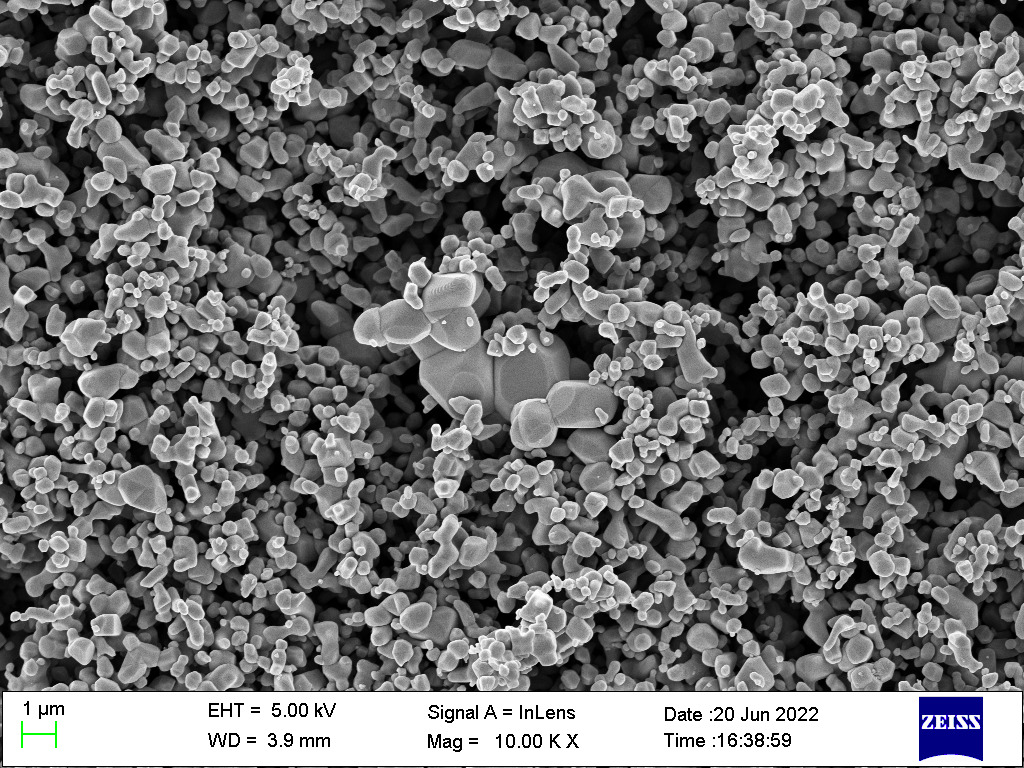}
    \caption{}
    \label{fig:SEM_TiO2}
     \end{subfigure}
     \hfill
     \centering
     \begin{subfigure}[b]{0.46\textwidth}
    \centering
    \includegraphics[width=85 mm]{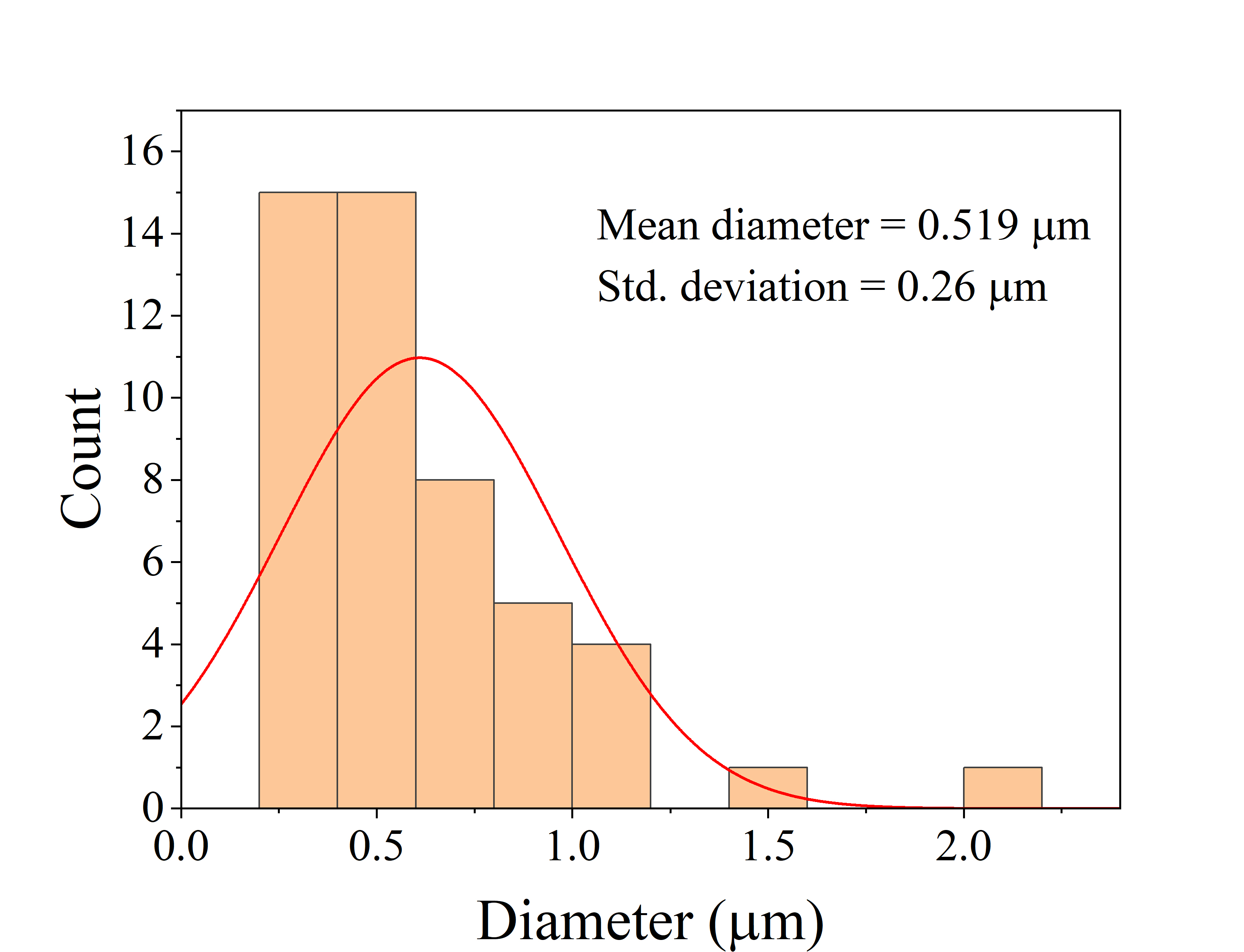}
    \caption{}
    \label{fig:size_distribution}
     \end{subfigure}
     \hfill
    \centering
    \begin{subfigure}[b]{0.46\textwidth}
    \centering
    \includegraphics[width=85 mm]{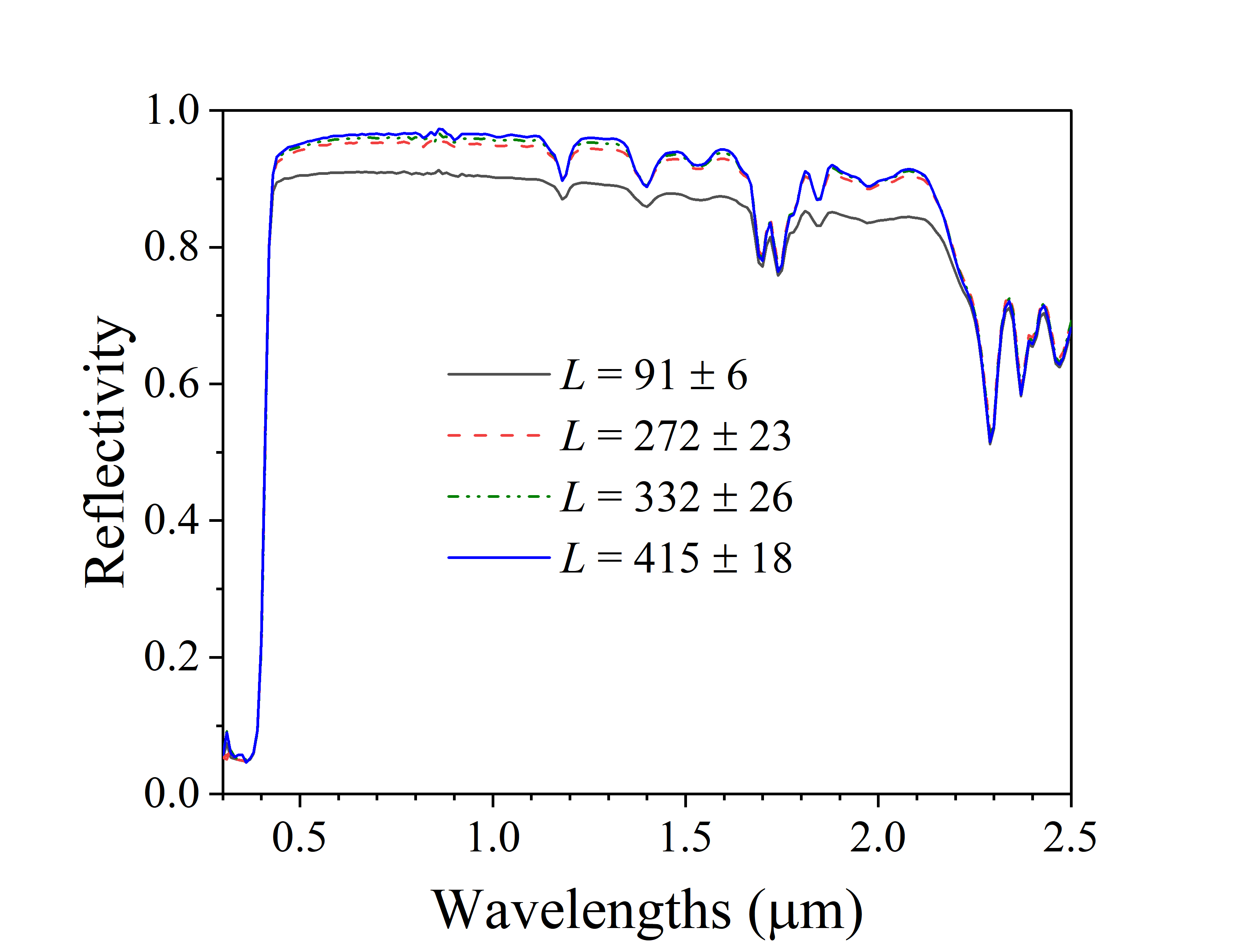}
    \caption{}
    \label{fig:R_for_diff_L}
     \end{subfigure}
    \hfill
    \centering
     \begin{subfigure}[b]{0.46\textwidth}
    \centering
    \includegraphics[width=85 mm]{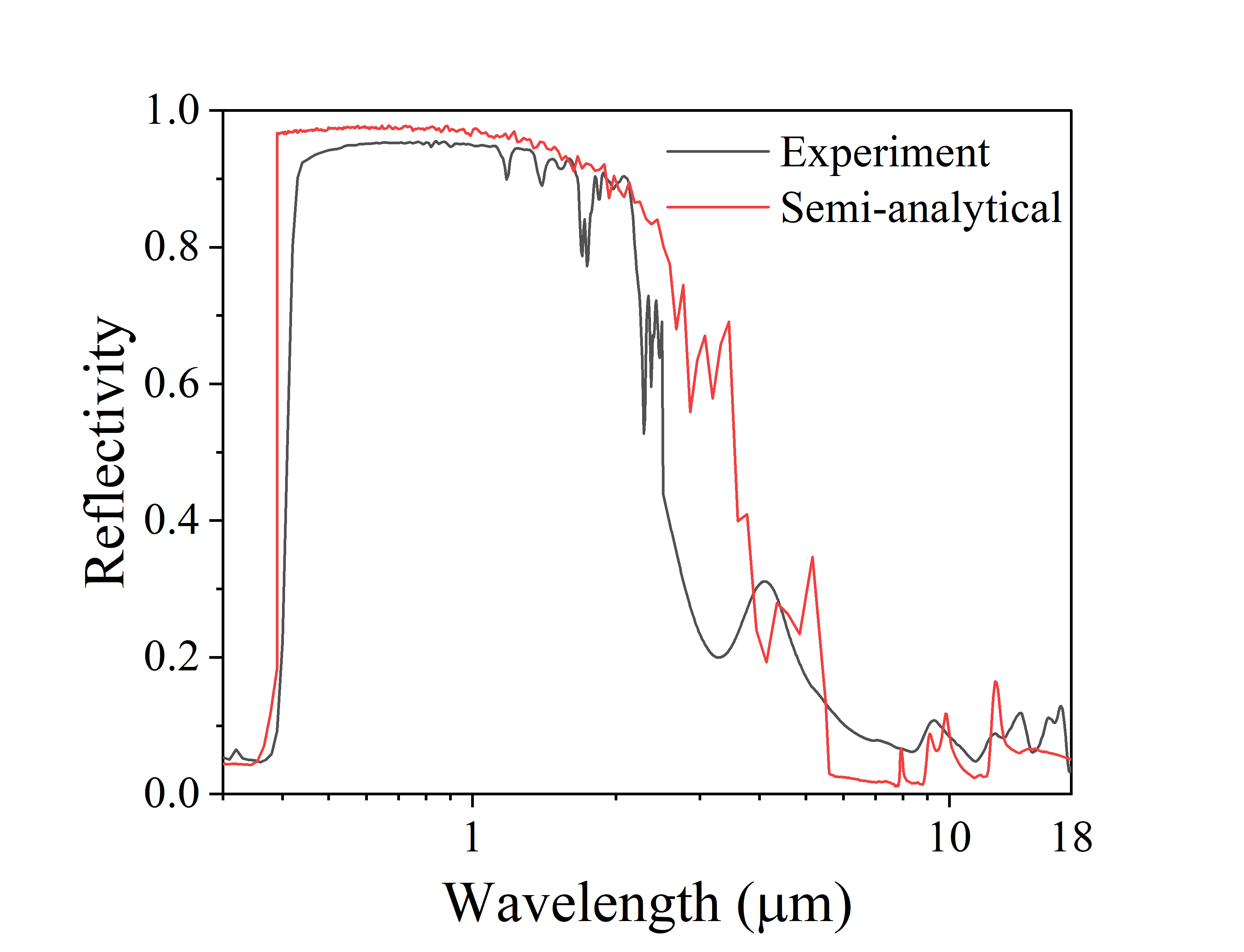}
    \caption{}
    \label{fig:R_exp_sim}
     \end{subfigure}
    \caption{(a) SEM image of TiO$_2$ pigment. (b) Size distribution of TiO$_2$ particles used to fabricate the paint coating. (c) Reflectivity of fabricated paint coating for various thicknesses. (d) The measured reflectivity of fabricated paint coating is compared with simulated reflectivity using semi-analytical technique.}
\end{figure}

We employed a field emission scanning electron microscope (Zeiss Ultra 55) to image and characterize the pigment powder particles, as shown in \Cref{fig:SEM_TiO2}. Particle size distribution was analyzed with ImageJ and is presented in \Cref{fig:size_distribution}. The pigment mean radius is 0.255~$\mu$m that matches with the optimized particle size of TiO$_2$. The size of the TiO$_2$ powder ranges from 0.1~$\mu$m to 1.5~$\mu$m. This size distribution provides high reflection in entire solar spectrum (smaller particles contribute in visible spectrum and larger particles in NIR) which can be seen in the reflectivity spectrum of the TiO$_2$/PDMS coating given in \Cref{fig:R_exp_sim}. Spectral reflectivity in the 0.3--2.5~$\mu$m wavelength range was obtained using  PerkinElmer Lambda 950 UV-Vis-NIR spectrometer with 150~mm integrating sphere, and in the 2.5-18$\mu$m range using PerkinElmer Frontier FTIR spectrometer with 75~mm integrating sphere. The reflectivity of the sample in 0.3--2.5~$\mu$m wavelength range was measured with respect to spectralon reflectance standard. The obtained relative reflectivity is then multiplied with the reflectivity of the spectralon reflectance standard (given in Figure~S5 in supplementary) to get the absolute reflectivity of the TiO$_2$/PDMS coating. The thickness of the coating was measured via a Fujitech\textsuperscript{TM} DFT gauge.
The reflectivity of our fabricated TiO$_2$/PDMS coating, with a thickness of 280~$\mu$m, is presented in \Cref{fig:R_exp_sim}. It aligns closely with the simulation results of our semi-analytical approach. Some discrepancies exist likely due to difference in the sizes of particle clusters captured in ImageJ, and breakdown of such clusters into smaller sized particles during the high-shear fabrication process. The spectral reflectivity of the TiO$_2$/PDMS coating for various thicknesses, shown in \Cref{fig:R_for_diff_L}, reinforces the increase in reflectance with increasing coating thickness. However, the rate of increase in reflectivity decreases significantly beyond a coating thickness of 250-300~$\mu$m.

The weighted solar reflectivity of the fabricated coating is 88.2~\% for the thickness of 280~$\mu$m. The maximum reflectivity of the paint is observed to be 89.3~\% for the thickness of the 415~$\mu$m. It is to be noted that reduction in reflectivity is due to absorbing nature of TiO$_2$ in UV region. In visible and NIR, the paint's reflectivity is greater than 94~\%. The emissivity (calculated using $\int I_{\text{BB}}(\lambda) \epsilon(\lambda) d \lambda / \int I_{\text{BB}}(\lambda) \, d\lambda$, where $I_{\text{BB}}$ is Planck's blackbody radiation) in the atmospheric transparency window (i.e. 8-13~$\mu$m) is seen to be 92.4~\%. The total emissivity of the coating for thicknesses 91, 272, 332, and 415~$\mu$m is also calculated using an emissometer (Devices \& Services, model AE1). The obtained emissivities for 91, 272, 332, and 415~$\mu$m thicknesses are 0.8, 0.85, 0.86, and 0.88, respectively.
.

\subsection{Cooling performance of the coating}

Many previously reported studies \cite{Mandal2018, Zhang2021Zro2, bijarniya2022experimentally, dasultra} assess the cooling performance of coatings on flat substrates, measuring temperatures at the substrate's back surface or between the coating and substrate. This approach doesn't mimic real-world scenarios where coatings are applied to structures to cool enclosed spaces.  In addition, this method may yield inaccurate temperature readings if one side of the thermocouple is exposed to ambient conditions or in contact with the substrate. To better simulate real-world conditions, we evaluate the cooling efficiency of the RC paint when applied on a hollow 4mm-thick aluminium box (7cm x 7cm x 7cm). The paint is applied to the box's exterior via spray gun or brush, and a top hole accommodates a thermocouple (sealed with cotton wool) to measure the internal temperature of the box.

\begin{figure}[t]
     \centering
     \begin{subfigure}[b]{1\textwidth}
    \centering
    \includegraphics[width=80 mm]{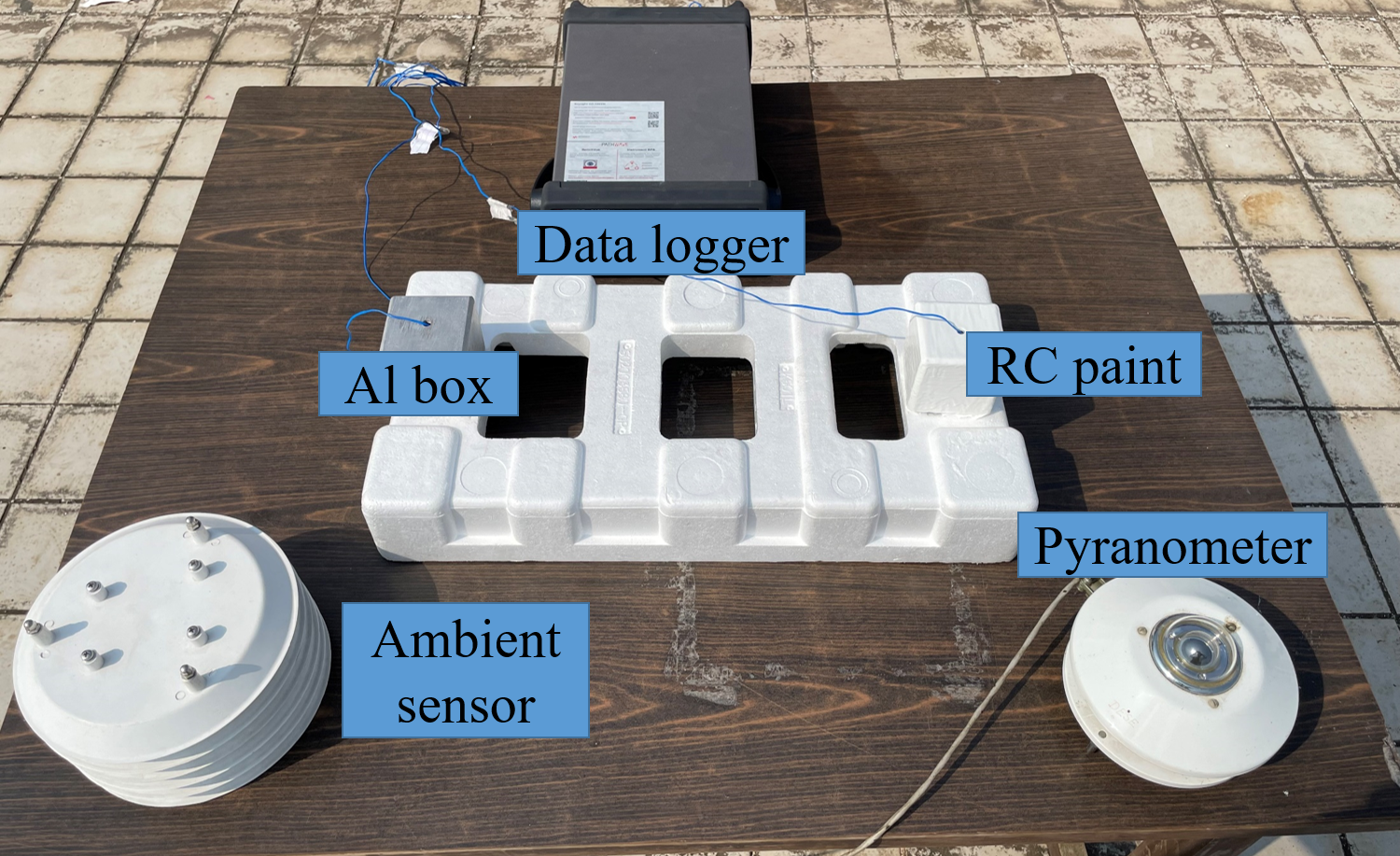}
    \caption{}
    \label{fig:exp_setup}
     \end{subfigure}
     \hfill
     \begin{subfigure}[b]{0.46\textwidth}
    \centering
    \includegraphics[width=85 mm]{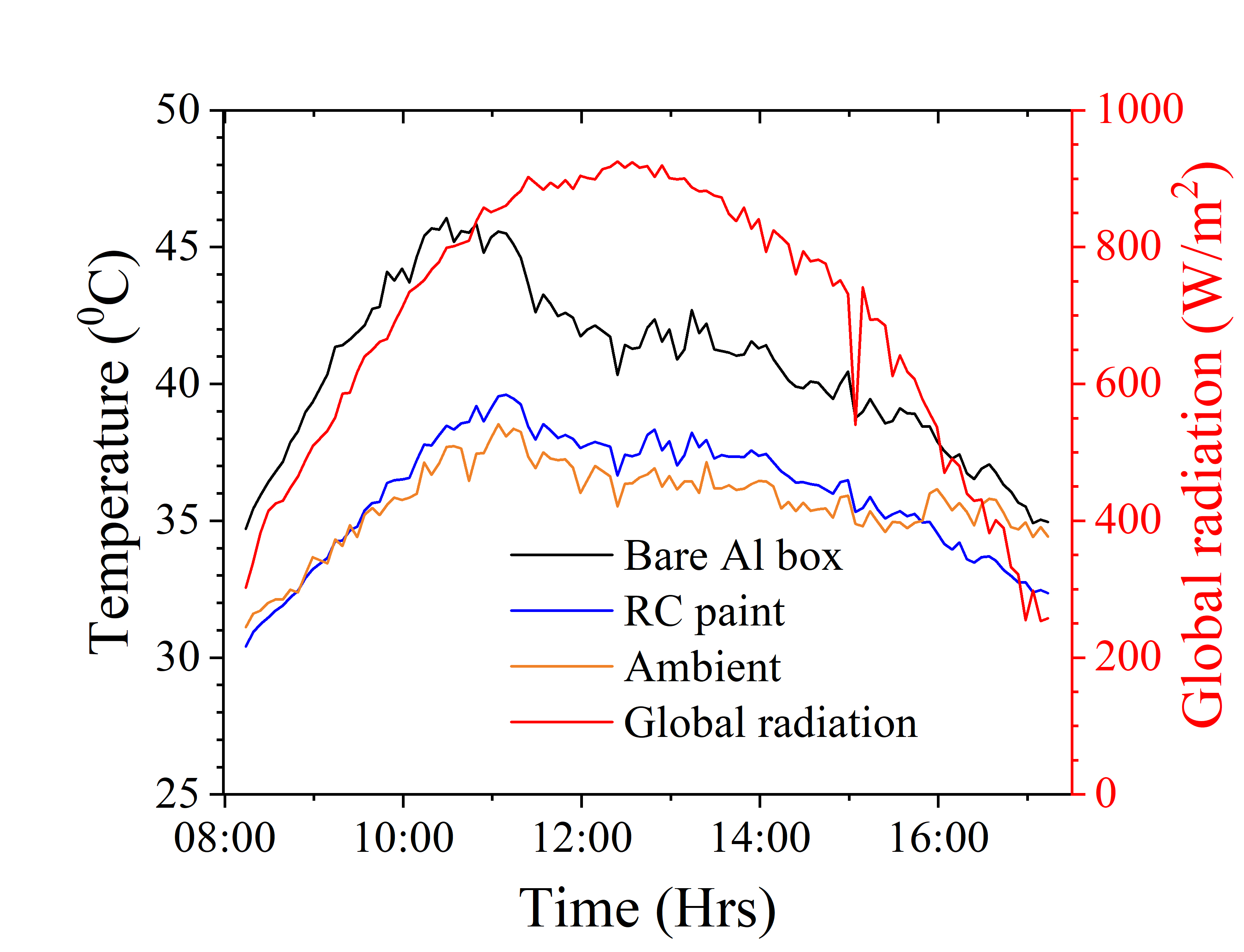}
    \caption{}
    \label{fig:25_april_boxes}
     \end{subfigure}
     \centering
     \hfill
     \begin{subfigure}[b]{0.46\textwidth}
    \centering
    \includegraphics[width=85 mm]{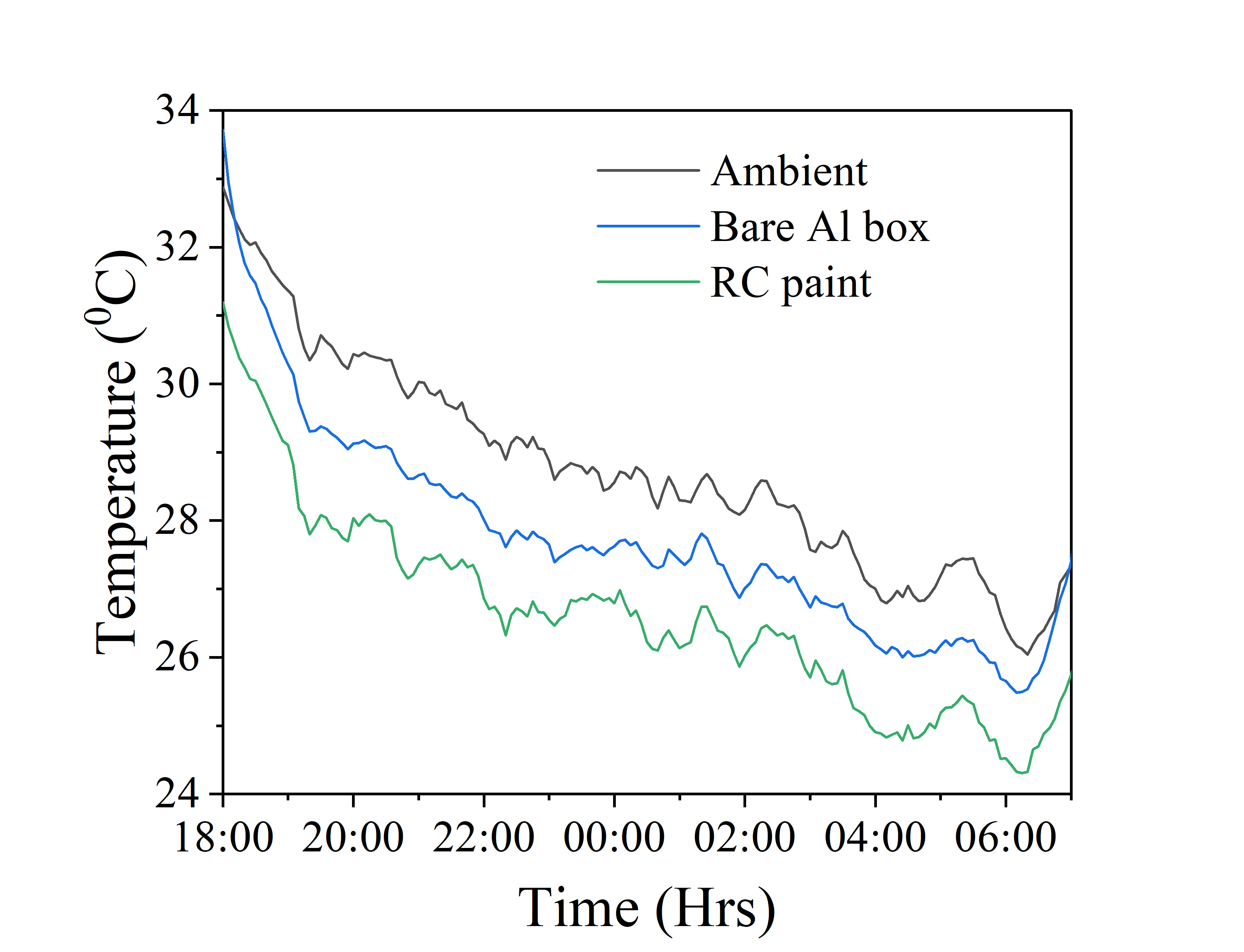}
    \caption{}
    \label{fig:nighttime}
     \end{subfigure}
    \caption{(a) Experimental setup to measure the cooling performance of RC coating. (b) Day time cooling performance of the RC paint measured on 25$^\text{th}$ April 2023. (c) Night time cooling performance of the RC coating measured on 15$^\text{th}$ April.  }
    \label{fig:cooling_performance}
\end{figure}

We used a Keysight DAQ970A data logger to record temperature data from T-type thermocouples at regular intervals. Each thermocouple was calibrated with a Julabo calibration bath. Global solar irradiation at the experimental site was measured using a Dynalab Weathertech pyranometer (model 147059). Ambient temperature was recorded by a radiation- and wind-shielded device from TrackSo, accurate to ±0.5°C (refer to \Cref{fig:exp_setup}). For the experiment, we employed two aluminum boxes: one coated with TiO$_2$/PDMS paint (labeled as RC paint in figures) and an uncoated one as a control (labeled as bare Al box in figures). All outdoor experiments were conducted on the rooftop of the Department of Energy Science and Engineering at the Indian Institute of Technology Bombay, Mumbai.

\Cref{fig:25_april_boxes} illustrates the cooling performance of the RC paint coated on the Al box, comparing it with temperatures from the baseline (bare Al box) and the ambient environment. All recorded temperatures rise in tandem with increasing solar irradiation, yet the increase is notably more pronounced in the baseline Al box. At 1019 hrs, the maximum cooling differential between the coated Al box and the baseline reaches 7.9~$^\circ$C. This significant difference is due to high reflectivity of the paint in visible and NIR.  During peak solar radiation hours, the RC paint-coated box registers temperatures exceeding ambient, but the peak difference is limited to 2.1~$^\circ$C at 1044 hrs. Sub-ambient cooling is evident post 1600 hrs (and also before 0900 hrs), but not observed during peak solar radiation hours. This behavior aligns with Mumbai's high relative humidity levels, which effectively "closes" the atmospheric transmission window, as further detailed in \cref{subsubsec:effect_humidity}.
In night-time conditions (\Cref{fig:nighttime}), both Al boxes — bare and RC paint-coated — record sub-ambient temperatures. The day-time cooling performance can be attributed to coating's high solar reflectivity while the night-time cooling points to the coating's emissive properties, even if potentially suppressed due to the high humidity conditions. Cooling results for the coating on substrates other than metallic are consistent with these findings (see Supplementary Figure S1 for data with wooden boxes).

\subsection{Potential reasons for suboptimal cooling performance}
Sub ambient cooling has been shown using TiO$_2$-particle based coating with similar reflectivity and emissivity in previous works \cite{Xue2020,song2022durable,du2022daytime}. However, we did not observe sub ambient cooling. It thus becomes important to understand what is causing the decreased cooling potential.

The cooling efficacy of a RC coating is influenced by atmospheric factors, notably humidity and air temperature. Particularly, atmospheric transmittance in the 8-13~$\mu$m wavelength is sensitive to these conditions. The design of outdoor experiments and their reporting methods are also crucial. The ensuing section delves into the challenges of achieving sub-ambient cooling and the standards in reporting radiative cooling results in prior literature.
 
\subsubsection{Ambiguity in the measurement of the ambient temperature}

\begin{figure}[t]
   \centering
  \includegraphics[width=85 mm]{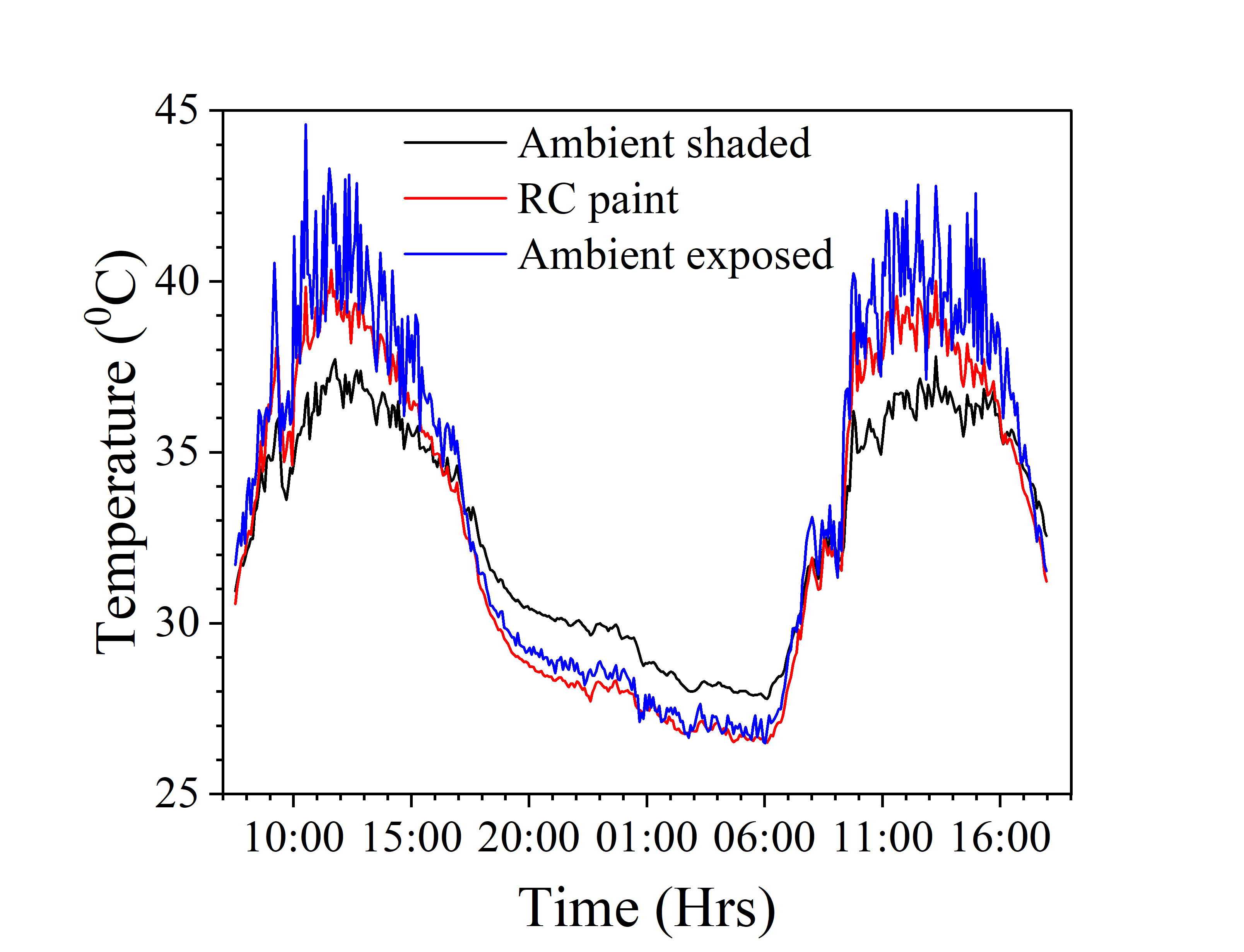}
   \caption{The comparison of the temperature of RC paint and ambient temperature. Here, ambient shaded is the ambient temperature measured by thermocouple kept inside a radiation and wind shield, whereas ambient exposed is the ambient temperature measured by a thermocouple exposed to sun and wind.}
  \label{fig:ambient_comp}
\end{figure}

The performance of RC coatings is typically assessed by comparing the temperature recorded behind the coating with ambient temperature. However, standard practices for measuring ambient temperature often involve using a thermocouple that's exposed to environmental elements like sun and wind \cite{Mandal2018, Wang2021, Zhang2021Zro2}, which can cause significant reading fluctuations and inaccuracies. A better alternative is using a radiation and wind shield, as shown in \Cref{fig:exp_setup}. This shield minimizes exposure to the sun and regulates airflow, enabling more precise and consistent measurements.
The importance of this improved measurement method is underscored by its impact on evaluating cooling performance. For instance, \Cref{fig:ambient_comp} shows the two ambient temperatures, one measured by a thermocouple placed inside a radiation shield and another exposed to sun and wind. The former measures lower ambient temperatures during the day. The opposite is observed at night, owing to cooler winds. This figure also presents the temperature profile for an RC paint-coated aluminium box. Measurements using an unshielded thermocouple would inaccurately suggest that the paint cools below ambient temperatures during the day, contradicting the shielded measurement. Accurate ambient temperature measurement is thus critical for reliable sub-ambient cooling reporting, and shielded devices are recommended for consistent RC coating performance evaluation.

\subsubsection{Using polyethylene sheet as a convection shield}

\begin{figure}[t]
     \centering
     \begin{subfigure}[b]{0.46\textwidth}
    \centering
    \includegraphics[width=80 mm]{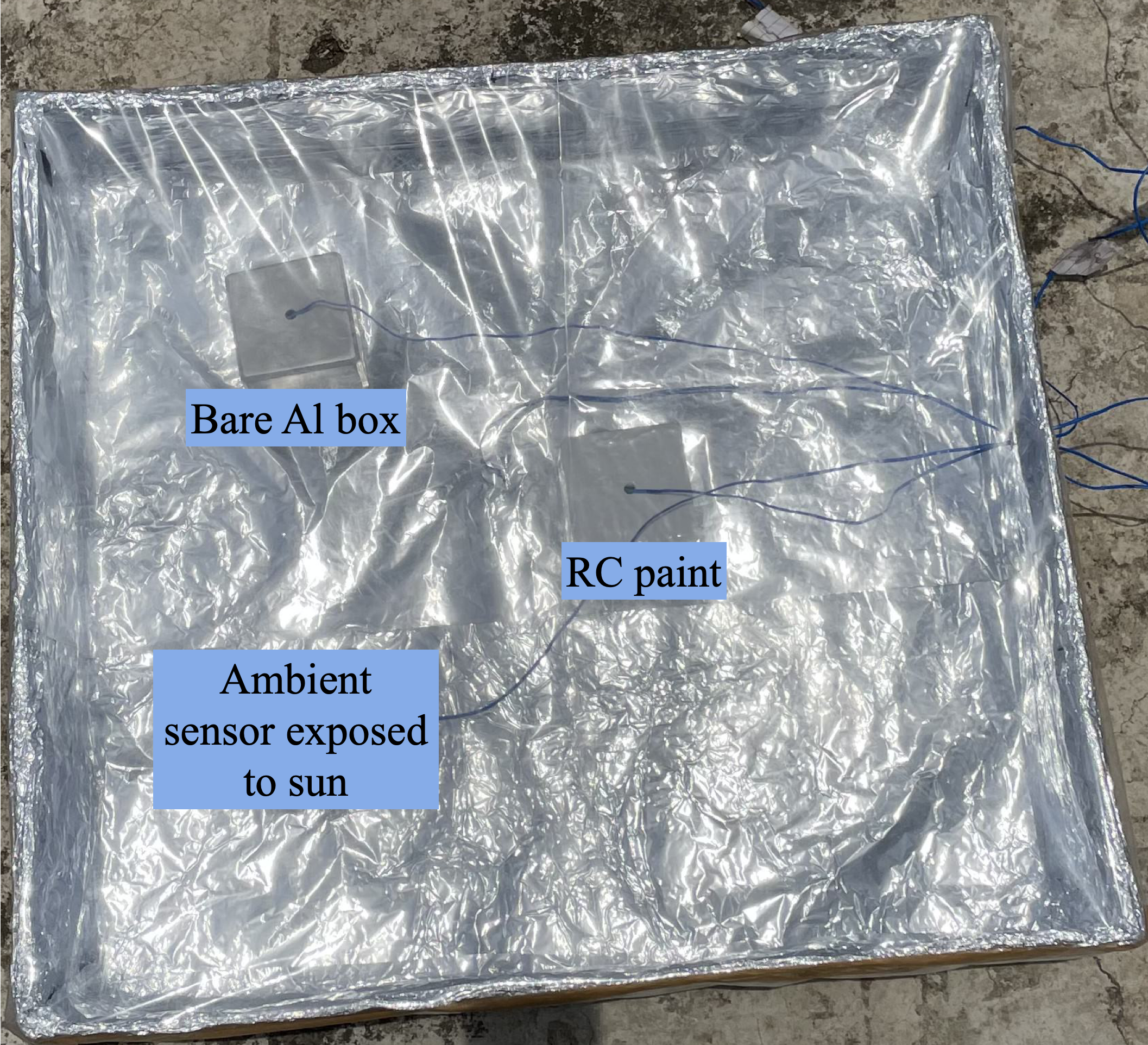}
    \caption{}
    \label{fig:setup_with_PE}
     \end{subfigure}
     \hfill
     \begin{subfigure}[b]{0.46\textwidth}
    \centering
    \includegraphics[width=85 mm]{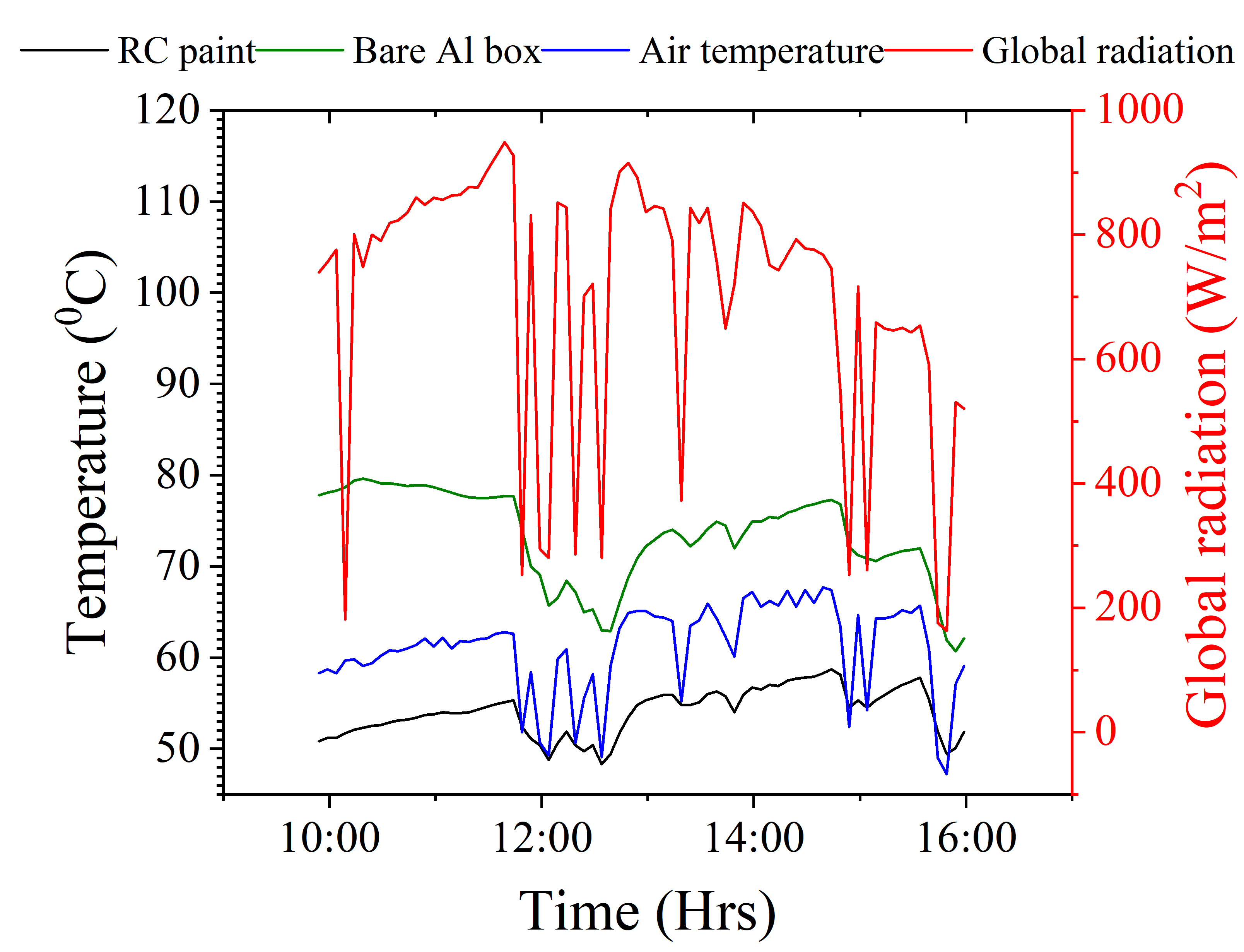}
    \caption{}
    \label{fig:PE_cover_temp}
     \end{subfigure}
     \caption{(a) Cooling performance testing setup with a radiation shield and PE cover. (b) The temperature of RC paint-coated Al box, bare Al box, and air temperature measured inside radiation shield box covered with PE on 11$~\text{th}$ June 2023.}
     \label{fig:effect_PE}
\end{figure}

The cooling performance of the RC coating is greatly affected by conduction and convection losses. To address this, studies report the use of a radiation shield with a polyethylene (PE) film covering on the top side to counter these parasitic losses \cite{Raman2014,Kou2017, Li2021baso4, Zhang2021Zro2, du2022daytime}. The radiation shield, which is typically made of aluminum foil, keeps the coating directly facing the sky, while the PE film controls convective heat loss. In an experiment to check the effect of such a setup, we measured temperatures of RC paint-coated Al box, a bare Al box, and air inside of this PE-covered setup as shown in \Cref{fig:setup_with_PE}. \Cref{fig:PE_cover_temp} shows a significant increase in air temperature and bare Al box, ultimately giving subambient cooling throughout the day with maximum cooling of 11.5~$^\circ$C below the ambient and 21.4~$^\circ$C compared to bare Al box.   
The PE film, while reducing convective heat loss, also trapped heat, raising the air temperature inside to 60-70°C, similar to a greenhouse effect. Hence, the use of PE films as convective shields is not recommended for measuring cooling performance, as they can significantly skew results.

\subsubsection{Effect of humidity}
\label{subsubsec:effect_humidity}

The atmospheric conditions of the local environment strongly impact the cooling performance of the RC coating. The increase in humidity and air temperature reduces the atmospheric transparency in 8-13~$\mu$m wavelength, which in turn reduces the ability of the coating to radiate heat to outer space via the atmospheric transparency window.
As a result, the cooling performance of a RC coating reduces significantly in highly humid areas in contrast to dry areas.  The effect of local humidity and air temperature on the cooling performance of RC coating has been studied by researchers for different locations \cite{liu2019effect,dong2019nighttime,bijarniya2020environmental,song2021temperature,huang2022effects}. Along similar lines, we seek to understand the role of humidity in affecting the performance of our RC coating in Mumbai where the experiments were carried out.   
For this, we calculate the atmospheric transparency using MODTRAN \cite{MODTRAN}. MODTRAN solves the atmospheric radiative transfer equation and is thus used to predict the optical properties of the atmosphere. It inputs precipitable water vapour (PWV) to calculate the atmospheric transmittance. The PWV (or total water column) is related to the relative humidity (RH) and ambient temperature. We adopt the method detailed in Ref.  \cite{dong2019nighttime} to calculate PWV from RH and ambient temperature (refer to Section S1 in Supplementary information for explanation about the procedure).

\begin{figure}[t]
     \centering
     \begin{subfigure}[b]{0.46\textwidth}
    \centering
    \includegraphics[width=85 mm]{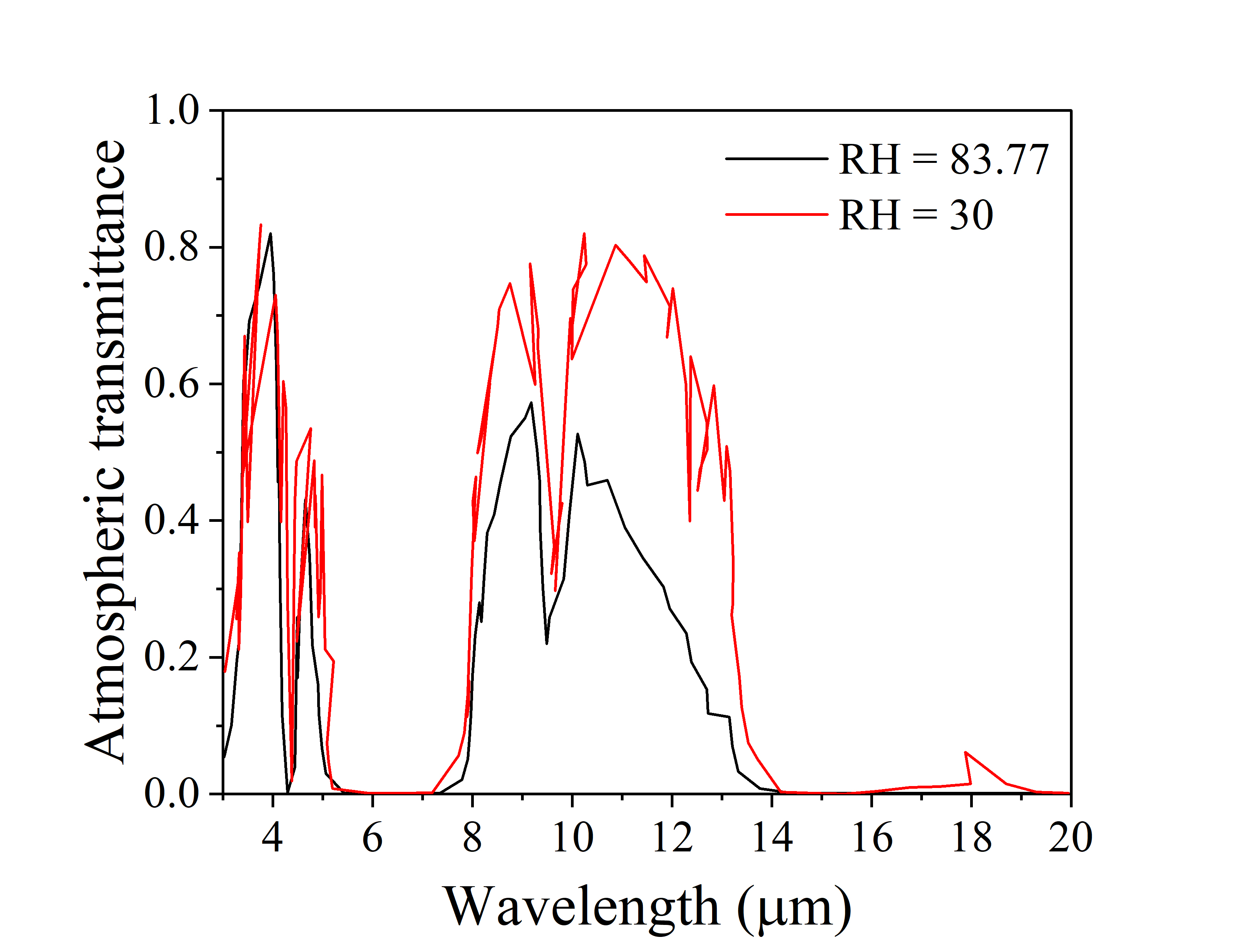}
    \caption{}
    \label{fig:ATW}
     \end{subfigure}
     \hfill
     \begin{subfigure}[b]{0.46\textwidth}
    \centering
    \includegraphics[width=85 mm]{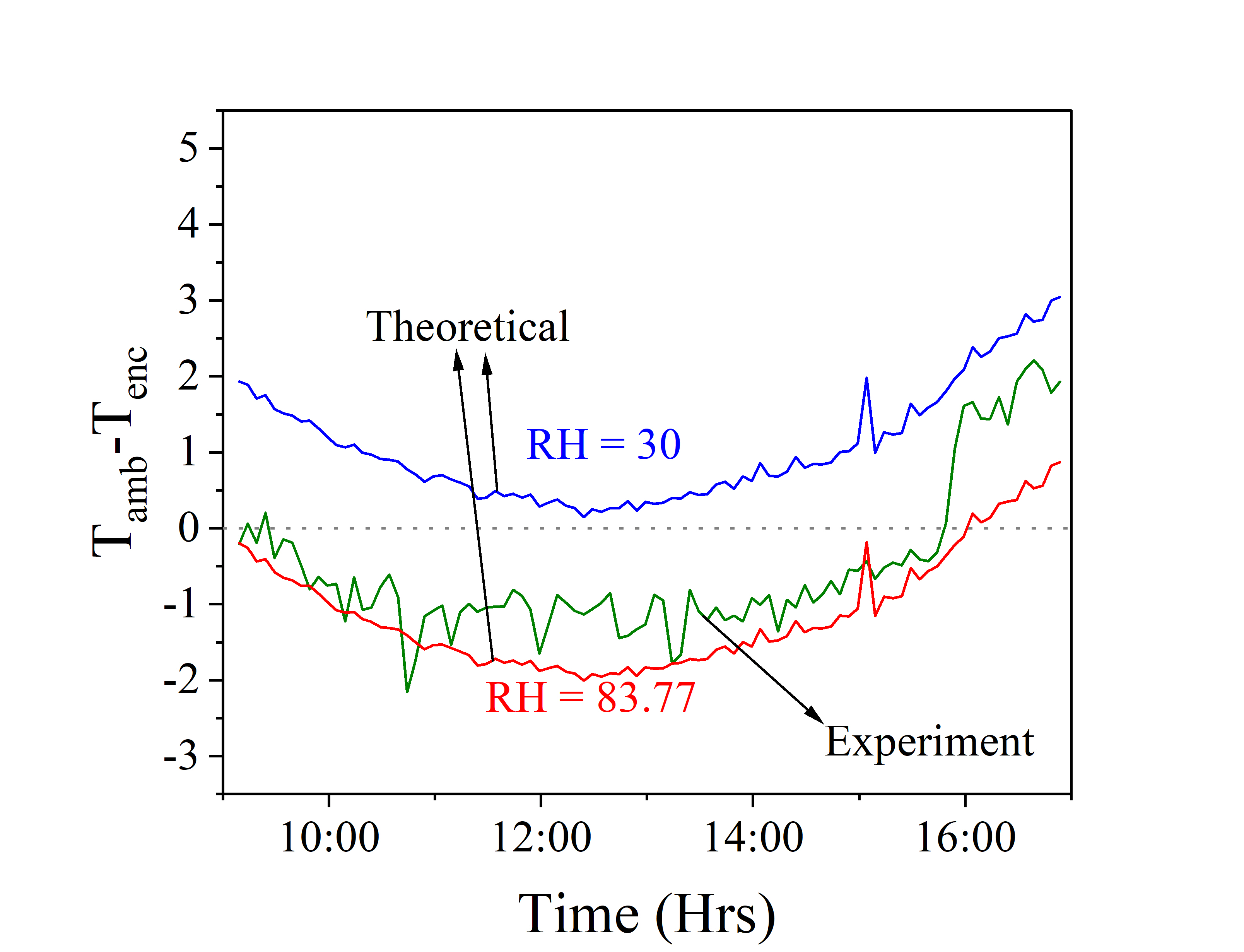}
    \caption{}
    \label{fig:Trad_exp_thr}
     \end{subfigure}
     \caption{(a) The atmospheric transmittance calculated using MODTRAN for two RH viz; 30~\% and 83.77~\%. (b) The comparison of differences of ambient and RC paint's measured and theoretically calculated temperature.}
     \label{fig:RH_effect}
\end{figure}

The calculations in this section are carried out for atmospheric parameters (83.77~\% average humidity and 29.55~$^\circ$C average ambient temperature) for 25$^\text{th}$ April 2023, obtained from the CPCB database \cite{cpcb}. 
\Cref{fig:ATW} shows the atmospheric transmittance calculated using MODTRAN at this ambient condition and the corresponding PWV value =~5.54~cm. 
These parameters are then utilized in the standard energy balance equation, as described in \cite{Raman2014,Mishra2021},  applicable to a radiative cooling system (refer to section S2 in the Supplementary information). Unlike these, and other studies, which focus on measuring the temperature of the coating, our work focuses on measuring the  more practical enclosure temperature. Consequently, the non-radiative heat loss in the energy balance equation needs more careful consideration. 
The non-radiative heat loss $P_{\text{cond+conv}}$ from the paint coating includes both losses to the ambient as well as to the Al box enclosure and can be quantified using:
\begin{equation}
   P_{\text{cond+conv}} = h_\text{o} ( T_\text{c} - T_{\text{amb}}) + h_\text{c-Al} ( T_\text{c} - T_{\text{enc}})
\end{equation}
where, the coating temperature, $T_\text{c}$, and interior temperature of the Al box, $T_\text{enc}$, can be related using
\begin{equation}
    h_\text{eff} ( T_\text{enc} - T_{\text{amb}})=h_\text{c-Al} ( T_\text{c} - T_{\text{enc}}).
\end{equation}
Here, $h_\text{o}$ (=10.43~Wm$^{-2}$K$^{-1}$) is the external convective heat transfer coefficient estimated using empirical relations \cite{test1981heat} based on the prevalent wind speed on that day, $T_{\text{amb}}$ is the ambient temperature, $h_\text{c-Al}$ is the effective heat transfer coefficient between coating surface and  still-air in the enclosure of the Al box, and $h_\text{eff}$ is the overall effective heat transfer coefficient of the setup from enclosure space to the ambient, which can be quantified using:
\begin{equation}
    \frac{1}{h_\text{eff}} = \frac{1}{h_\text{o}} + \frac{l_\text{c}}{k_\text{c}} + \frac{1}{h_\text{c-Al}} ;
\end{equation}
and
\begin{equation}
    \frac{1}{h_\text{c-Al}} = \frac{l_\text{Al}}{k_\text{Al}} + \text{TBR} +\frac{1}{h_\text{i}} ;
\end{equation}

Here, $l_\text{c}$ and $k_\text{c}$ represent the thickness and thermal conductivity of the RC paint, while $l_\text{Al}= 4~$mm and $k_\text{Al}= 237~$Wm$^{-1}$ K$^{-1}$ are those of the Al wall, with $h_\text{i}$ denoting internal heat transfer between still air and the Al wall. TBR, which is the thermal barrier resistance between polymer coating and Al surface \cite{fuller2001thermal,bendada2004analysis}, and the thermal resistance due to the Al wall, are small compared to $1/h_i$ and are neglected.  A standard value of 2.6~Wm$^{-2}$K$^{-1}$ is used for internal heat transfer coefficient \cite{qengel2014heat, gerlich2011modelling}. The value $k_c$ is obtained from Maxwell's relation \cite{Maxwell1873} $k_\text{c} = k_\text{m} (1+3f/((k_\text{p}+2k_\text{m})/(k_\text{p}-k_\text{m})-f$, where subscript m stands for matrix and p stands for particle. The thermal conductivity of PDMS and TiO$_2$ are taken to be $k_\text{m} = 0.2~$Wm$^{-1}$K$^{-1}$ and $k_\text{p} =4.8~$Wm$^{-1}$K$^{-1}$ respectively.

Using the atmospheric transmittance from \Cref{fig:ATW} and with conduction and convection heat loss as mentioned above, we can calculate the equilibrium temperature $T_{\text{enc}}$ of the RC paint coated Al box by solving the energy balance equation (refer to section S2 in Supplementary information).  
The deviation of  $T_{\text{enc}}$ from measured ambient temperature, which is an indicator of cooling performance of the RC coating, is plotted in \Cref{fig:Trad_exp_thr} and compared with experimental measurements. An appreciable  match with the experimental measurements is observed.
\Cref{fig:Trad_exp_thr} also contrasts the cooling performance under a less humid condition (RH = 30~\%). Other parameters, such as ambient temperature and heat transfer coefficient, are left untouched.  A marked decrease in the cooling performance is noticed in \Cref{fig:Trad_exp_thr} with increased humidity levels. Consequently, RC coatings deployed in highly humid regions will experience an appreciable drop in their emissive performance, translating to difficulties in achieving sub-ambient cooling.

We have shown conclusively that uncontrollable external factors, including humidity, ambient temperatures, and cloud cover influence the cooling efficacy of the paint. Since the goal of developing such material coatings is to minimize reliance on energy-consuming cooling systems like air conditioners, achieving sub-ambient temperatures should not be the sole criterion for evaluating RC coating efficiency. Instead, temperature reductions recorded in outdoor tests, which consider the prevailing atmospheric conditions, offer a more relevant benchmark. The methodology employed here, utilizing MODTRAN, facilitates the prediction of paint performance across various environmental scenarios. This enables a direct comparison of any newly developed coating's performance in differing climatic conditions.

\section{Conclusion}
This study introduces a low particle volume fraction  TiO$_2$/PDMS paint-coating designed for passive daytime radiative cooling. The coating, with 88.2\% solar reflectivity, and 92.4\% emissivity in the atmospheric transparency window, was tested outdoors in Mumbai, India. Unlike other studies that compare coating surface temperature to ambient, this research measured the enclosure space temperature which will be of more of practical utility in real-world applications. Though the coating is observed to significantly reduce the heat load in the enclosure space of a hollow aluminum box compared to an uncoated control, sub-ambient cooling was not achieved, largely due to Mumbai's high humidity which limits the effectiveness of the atmospheric transparency window. The study underscores that PDRC coatings could achieve sub-ambient cooling in low humidity conditions and notes the coating's considerable performance under existing conditions, reducing temperature by about 8°C. 
The study has also highlighted techniques currently adopted  to compare and report cooling performance vis-\`a-vis the measurement of ambient temperature, and the use of polyethylene sheet as a convective shield and strongly suggests better alternatives for the standardisation and ease of comparison of all future measurements. 
\section*{Supporting Information}
Supporting  Information  is  available  from the author.
\section*{Acknowledgments}
B.R.M. and S.S. acknowledge support from Prime Minister’s Research Fellowship (PMRF). K.S. acknowledges support from La Fondation Dassault Systèmes and SERB Grant No. SRG/2020/001 511. Authors thank P.P. Joshi from Thermogreen Cool Coat Pvt. Ltd. for his valuable guidance in paint development. The authors are also thankful to Dr. H.C. Barshilia from CSIR-National Aerospace Laboratories, Bangalore, India from whom the results from FTIR setup were obtained.
\section*{Conflict of Interests}
The authors declare no conflicts of interest.
\section*{Data Availability Statement}
Data underlying the results presented in this paper are not publicly available at this time but may be obtained from the authors upon reasonable request.

\bibliographystyle{WileyNJD-ACS}
\bibliography{ref}
\end{document}